\documentclass[aps,prb,showpacs,superscriptaddress,amsmath,amssymb,twocolumn]{revtex4}

\usepackage{mathptmx} 
\usepackage{helvet} 
\usepackage{courier} 
\usepackage{type1cm} 
\usepackage{makeidx} 
\usepackage{graphicx} 
\usepackage{epstopdf} 
\usepackage{verbatim} 
\usepackage{amsmath}
\usepackage{calc}
\usepackage{enumitem}
\usepackage{amssymb}
\usepackage{latexsym}
\usepackage{bm}
\usepackage{xcolor}

\makeindex 

\usepackage[
	colorlinks,
	linkcolor=blue,
	anchorcolor=blue,
	citecolor=blue,
	urlcolor=blue,
	menucolor=blue,
	filecolor=blue
]{hyperref}
\usepackage[capitalize]{cleveref}
\usepackage{xr}
\begin{document}

\title{Paradoxical creation of a polydomain pattern by electric field in BaTiO$_3$ crystal}

\author{Petr S. Bednyakov}
\affiliation{FZU-Institute of Physics of the Czech Academy of Sciences$,$ Na Slovance 2$,$ 18221 Praha 8$,$ Czech Republic}

\author{Petr V. Yudin}
\affiliation{FZU-Institute of Physics of the Czech Academy of Sciences$,$ Na Slovance 2$,$ 18221 Praha 8$,$ Czech Republic}
\affiliation{Kutateladze Institute of Thermophysics$,$ Siberian Branch of Russian Academy of Science$,$ Lavrent’eva av. 1$,$ Novosibirsk$,$ Russia}

\author{Alexander K. Tagantsev}
\affiliation{Swiss Federal Institute of Technology (EPFL), CH-1015 Lausanne, Switzerland}

\author{Ji\v{r}\'{\i} Hlinka}
\affiliation{FZU-Institute of Physics of the Czech Academy of Sciences$,$ Na Slovance 2$,$ 18221 Praha 8$,$ Czech Republic}

\date{\today}

\begin{abstract}
It is known that ferroelectric single crystals can be turned from a polydomain to a monodomain state by the application of an electric field.
Here we report an unexpected opposite effect: the formation of through-the-crystal polydomain pattern in a monodomain BaTiO$_3$ crystal in response to the applied electric field favoring the initial orientation of the polarization.
The effect is achieved for special electric field direction which equally selects two domain states, which are present in the polydomain pattern.
At the formation of the pattern,  the new wedge domains propagate from the sides of the sample in the direction transverse to the electric field.
The observations are rationalized in terms of a simple analytical model treating energies of competing domain configurations as functions of the electric field.
The results of the analytical treatment are supported by phase field modeling.

\end{abstract}

\maketitle

\section{Introduction}

The energy degeneracy with respect to the orientation of the order parameter is a basic feature of ordered solids like magnetics,  ferroelectrics, and ferroelastics \cite{cross1990ferroic}.
Due to this, such materials can exist in both single-domain  and multi-domain state.
In practice, each of  these states may be the required one depending on the application \cite{liu2022controllable,salje2009domain,shur2015micro}.
Thus, the way of swooping between them is an issue of appreciable interest \cite{scott2016review}.
A common way of bring a multi-domain  state to a single-domain one is the  application of a field conjugated to the order parameter (magnetic field, electric field, or mechanical stress) \cite{lai2006electric,chen2013electric}. 
We have identified a paradoxical situation where the application of the conjugated field to a basically single-domain sample brings it to a multi-domain state occupying the whole sample.

In this paper, we experimentally demonstrate that phenomenon within a tetragonal BaTiO$_3$ single crystal shaped into a bar elongated in the direction perpendicular to an electric field applied along the [110] direction.
To rationalize our observation, we formulated a scenario based on an analytical model and numerical simulations.

It is known that the formation of domain patterns in ferroelectrics is driven by an interplay between the domain wall energy and the energy of the depolarizing field \cite{mitsui1953domain}.
In our system due to finite size effects and the geometry of the sample, such an interplay becomes very specific, which makes  possible the field-induced  generation of the through-the-crystal domain pattern.

\section{Experimental} \label{ThM}

\begin{figure}[t]
\centering
\includegraphics{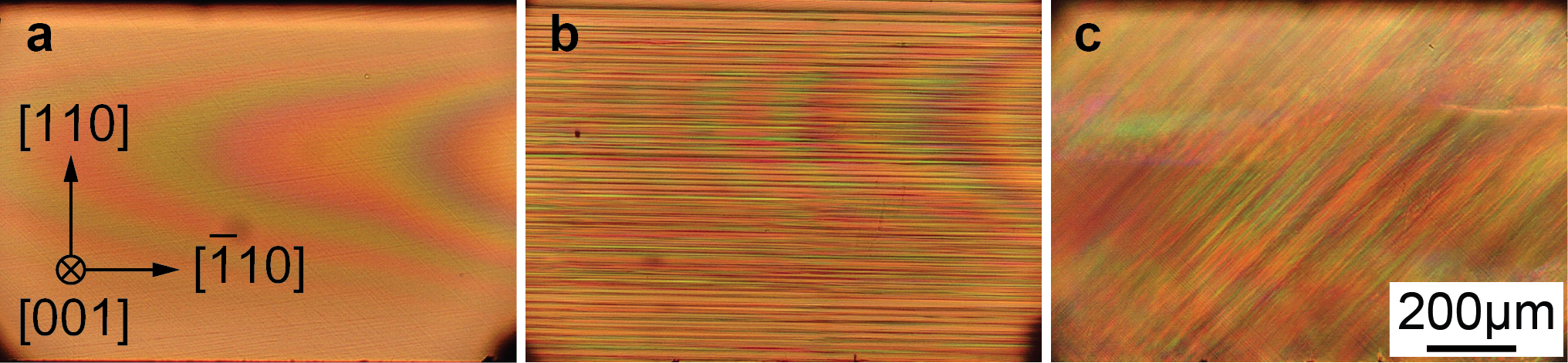}
\caption{
Three domain states observed in our BaTiO$_3$ sample: monodomain (a), with (110) domain walls (b) and with (101) domain walls (c).
The crystallographic orientation  of the sample is shown in panel (a).
The colored contrast is associated with the mechanical strain resulting from the application of the electric field ($<$0.5~kV/cm).
}
\label{DomainStructures}
\end{figure}

\begin{figure}[t]
\centering
\includegraphics{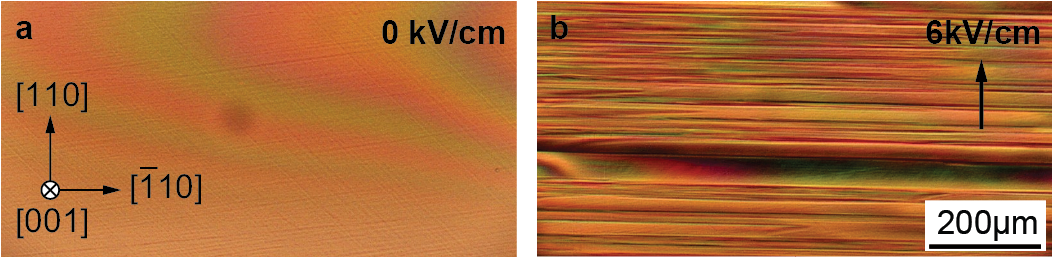}
\caption{
The initial monodomain state before poling (a) and the final polydomain state after a field of 6~kV/cm being  along [110] direction (b) observed in transmitted polarized light.
The crystallographic orientation of the sample is shown in panel (a).
}
\label{1st-last}
\end{figure}

\begin{figure}[t]
\centering
\includegraphics{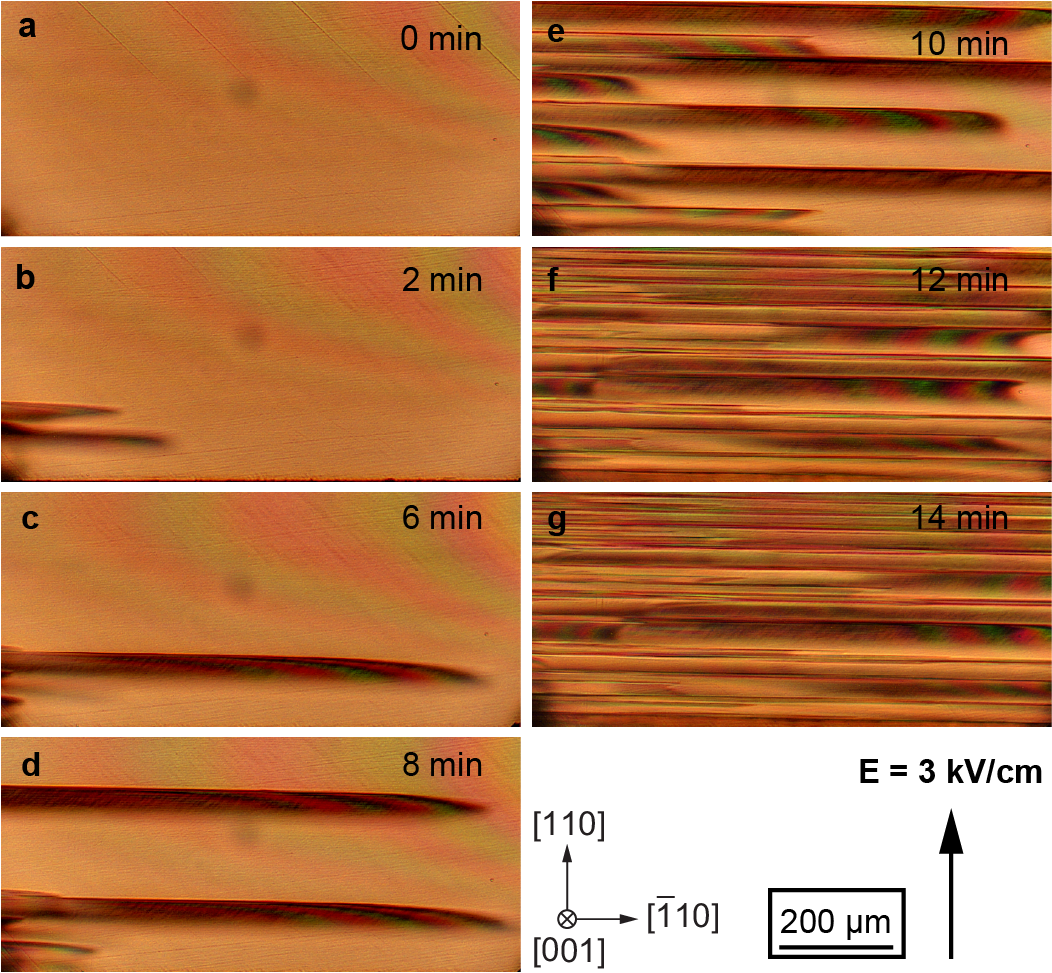}
\caption{
The time evolution of the domain structure in the BaTiO$_3$ sample under a field of 3~kV/cm.
The crystallographic orientation of the sample is shown in the bottom right panel.
}
\label{Evolution}
\end{figure}

In the tetragonal phase of barium titanate crystals, the domain structure comprises six possible domain states, with the spontaneous polarization parallel and antiparallel to the principal crystallographic axes.
In our experiment we used BaTiO$_3$ crystal having a shape of cuboid with the dimensions $1\times1\times5$ mm$^3$ obtained through the top-seeded solution growth technique. Two $1\times5$ mm$^2$ opposite faces of the sample, which are normal to  the [110] direction, are  electroded.
Thus an electric bias applied to the electrodes equally favors states with the spontaneous polarization parallel to [100] and [010].
We observed ferroelastic boundaries between them at room temperature using a polarizing microscope in transmission mode (see all the details in Appendix, section \ref{Methods}).
Typical domain configurations encountered under a weak  electric field ($<$0.5~kV/cm - applied voltage $V$~$<$~50~V) are presented in Fig.
\ref{DomainStructures}: a single domain state (a), a state with (110) domain walls (b), and a state with (101) domain walls (c).
We found that starting from a single-domain state, the application of an electric field of 6~kV/cm ($V$~=~600 V) for a few seconds along the [110] direction leads to the formation of a through-the-crystal polydomain pattern  with [100] and [010] domain states separated by (110) neutral domain walls (Fig.\,\ref{1st-last}).
We also found that such a polydomain pattern forms as well in the field  just exceeding  3~kV/cm ($V$~=~300 V).
However, the formation of the final state takes more time. The domain state evolution under such a field is illustrated in Fig.\,\ref{Evolution}.

Concisely, our observations are summarized as follows. At
electric fields ranging from 3 kV/cm to 6 kV/cm, we observed the appearance of ferroelastic wedge domains growing from the sides of the crystal. 
After such a poling procedure, we obtained a polydomain structure consisting of (110) domain walls.

The growth of the wedges at an applied electric field of 3~kV/cm progressed over time (Fig.\,\ref{Evolution} c-h).
We found that this process accelerates with increasing poling field.
They propagated from one side of the crystal to the opposite side and stabilized when they either reached opposite side of the sample, or when encountering similar wedges arriving from the opposite side or some defects.
The growth of these domain wedges stopped after reaching some critical domain wall density, with an average domain width of about 10~-~20~\textmu m.
Eventually, these domain wedges filled the entire volume of the monodomain part (Fig.\,\ref{Evolution}).

When the electric field along [110] direction was released, some of the domain wedges disappeared while through-the-crystal domain walls remained in the structure for months, as far as we could check. 
\section{Theoretical model} \label{ThM}

\begin{figure}[t]
\centering
\includegraphics{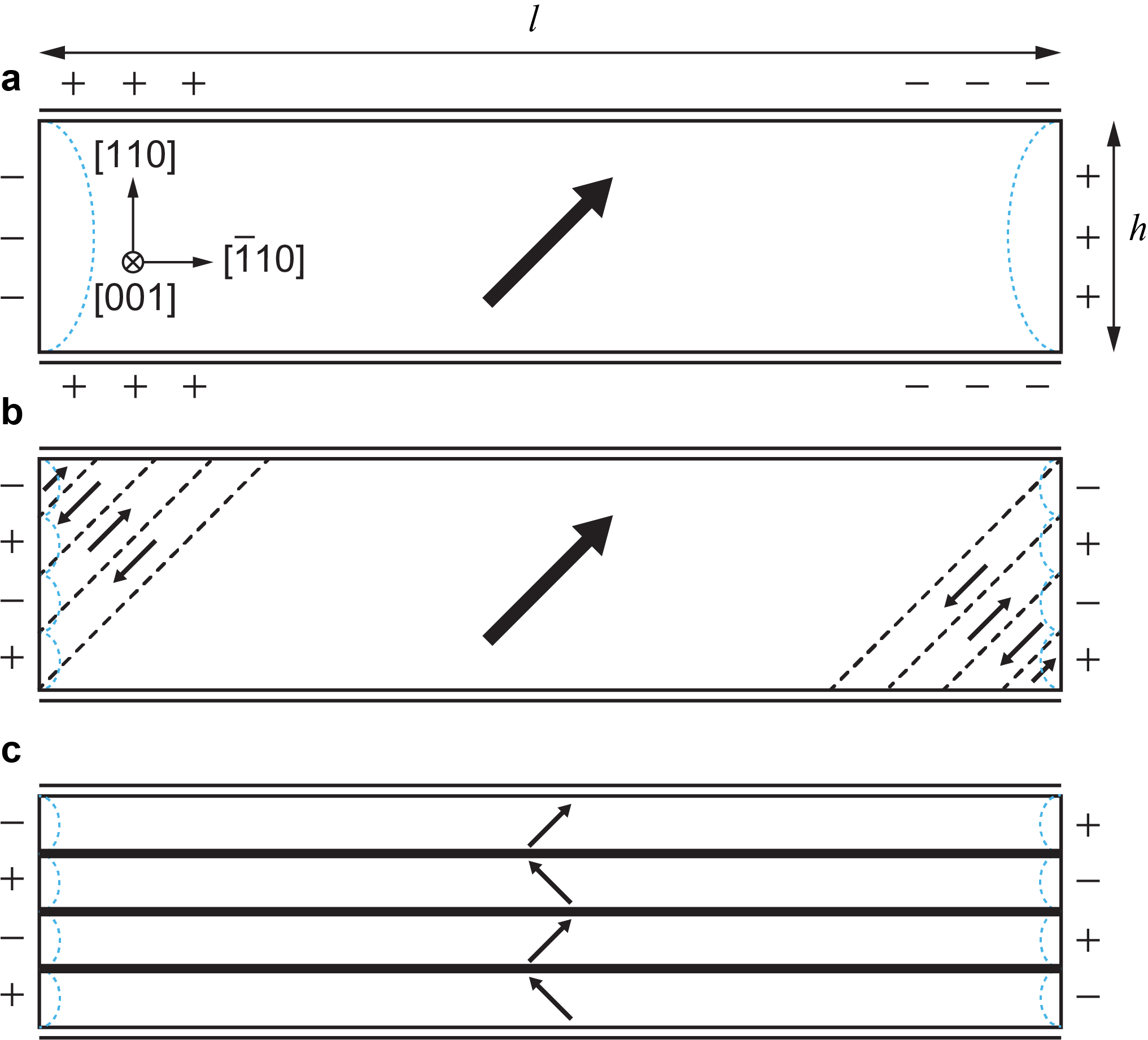}
\caption{
Three stable/meta-stable states of possible domain configuration  of tetragonal BaTiO$_3$ beam  - (a), (b), and (c).
The crystallographic orientation of the beam is shown in (a).
Thick lines and dashed lines show  $90^\circ$ and $180^\circ$ domain walls, respectively.
Electrodes are shown with double lines. 
Arrows show the orientation of the spontaneous polarization.
The sign of the total (bound + free) surface charge is indicated.
The regions with strong depolarizing field are schematically outlined with thin blue dashed lines. 
}
\label{Theory}
\end{figure}

For the interpretation of the experimental results it was important to take care of the fact that only ferroelastic $90^\circ$ domain walls are seen in the microscopic images, while non-ferroelastic $180^\circ$ domain walls may be present in a sample which looks homogeneous\cite{beccard2022nanoscale}.
In this regard, we were forced to consider scenarios where some actors are backstage.
Working through different scenarios and analyzing them led us to understand how the paradoxical growth of domain walls in the electric field can occur.
Here we give a simple model, which will be further validated by phase field numerical calculations.
In this model, the formation of such a pattern is addressed  in terms of three stable/meta-stable configurations: a single-domain state, a quasi-monodomain state where domain walls are present only near the lateral sides of the sample, and a through pattern of domain walls.
These states are schematically depicted in Fig.\,\ref{Theory}~a,~b, and c, where a side view of a beam with a shape of a cuboid is shown.
We will also shorthandly refer to these configurations as states (a), (b) and (c), respectively. 

It is known that in an electrostatically isolated thin ferroelectric plate, where an appreciable  non-zero component  of the spontaneous polarization is normal to it, the single-domain state is absolutely unstable because of the depolarizing effects.
This usually results in a multi-domain state (see Appendix, section \ref{Mitsui} and Fig.\,\ref{TheorySup} therein).
A simple theory accounting for this phenomenon was given by Mitsui and Furuichi \cite{mitsui1953domain} (hereafter, we will refer to this as Mitsui's theory).
Briefly, Mitsui's theory describes the partitioning of the system into domains, and finds the equilibrium distance between the domain walls from the balance between the energy of depolarizing fields and the energy of domain walls.
Here we generalize Mitsui's theory to describe all the three given states with one common formula.
Let us start with the single-domain state (a) where the energy of domain walls is absent and the only energy is electrostatic. 

Unlike in the case of a plate without electrodes, in the considered beam of the tetragonal BaTiO$_3$  with short-circuited electrodes  (Fig.\,\ref{Theory}~a), the single-domain state is thermodynamically stable.
Here the depolarizing field due to the bound charge of the $[110]$ component of spontaneous polarization is screened by the free charge exchanged between the electrodes.
At the same time, the field of the bound charge at the side surfaces  is compensated by an additional free charge at the electrodes such that  the depolarizing field becomes confined to a distance $\simeq h$ near the sides, where $h$ is a thickness of the crystal.
We assume that the length of the beam $l$ is large enough for the two regions with the strong depolarizing field to not overlap.
In the frame of our simplified theory, the energy of this state is proportional to the area occupied by the depolarizing field and amounts to (see Appendix, Fig.\,\ref{TheorySup}) 
\begin{equation}
\label{Ua}
U_a \approx 0.4 U_0h^2,
\end{equation} 
where
\begin{equation}
\label{U0}
U_0=\frac{0.3 P_0^2}{\chi}
\end{equation} 
is a characteristic energy density of the depolarising field; $P_0$ is the absolute value of spontaneous polarization; $\chi$ is the dielectric susceptibility which we assume to be isotropic. 
Here and throughout, we denote by $U$ the energy per unit length in the [100] direction.

Despite the fact that, in the system addressed, the single-domain is thermodynamically stable (with respect to a homogeneous change of the polarization vector value),  it may be unstable with respect to splitting into domains.
As we show in the Appendix for the states shown in Fig.\,\ref{Theory}~b,~c  the energy of the depolarizing field reduces to 
\begin{equation}
\label{UD}
U_D \approx 0.4 U_0hW,
\end{equation} 
where $W$ is the width of the band with constant sign of the bound charge on the sample side. 
For the state (b) $W=w/\sqrt{2}$ and for the state (c) $W=w$, where $w$ is the spacing between domain walls. 
The domain pattern is controlled by the trade-off between depolarizing field and domain wall energy.
The latter is inverse proportional to the spacing between domain walls:
\begin{equation}
\label{Us0}
U_w \approx \frac{\sigma h }{w}L,
\end{equation} 
where $L=h/\sqrt{2}$ for the state (b) and $L=l$ for the state (c) - the effective domain wall length, $\sigma $ is the wall energy per unit area, which we evaluate using a simple expression in terms of the wall thickness $t_W$ \cite{tagantsev2010domains}:
\begin{equation}
\label{sigma}
\sigma = \frac{P_0^2 t_W}{6 \chi}.
\end{equation} 
Minimizing the total energy of the system with respect to $W$  and taking into account (\ref{sigma}), we arrive at the following expression for the optimized energy of the states (b) and (c):
\begin{equation}
\label{Ub}
U_b \approx 0.6 U_0\sqrt{ht_W}h;
\end{equation} 
\begin{equation}
\label{Uc}
U_c \approx U_0\sqrt{lt_W}h.
\end{equation} 

Comparing Eqs.~(\ref{Ua}), (\ref{Ub}), and (\ref{Uc}) one 
checks that the energy of the single-domain configuration by many orders of magnitude exceeds that of the other configurations. It follows from the fact that, for any realistic values of the parameters of the problem, the 
inequality $t_Wl\ll h^2$ is very strong.

Comparing the energies of the other two configurations one sees that the energy of the state (c) increases with elongation of the sample.
Thus in samples with sufficient value of the aspect ratio $l/h$, before the application of the field, the system is expected to be in the state (b).
However, the application of the electric voltage $V$, which favors the orientation of polarization in the bulk of the beam, may change the energy balance.
In particular, state (b) stands out by containing a region of area $h^2/2$ in which the polarization is projected in the direction opposite to the applied electric field.
Under the action of the applied voltage, the energy (strictly speaking, the thermodynamic potential for a fixed potential at the electrodes) of the state (b) increases by the value of
\begin{equation}
\label{UE}
U_V = P_0 V h/\sqrt{2}.
\end{equation} 
Thus, above the critical voltage $V_f$ satisfying condition
\begin{equation}
\label{condc}
U_c = U_b+U_V,
\end{equation} 
the configuration with the $90^\circ$ walls (state (c)) becomes more favorable than that with the $180^\circ$ walls (state (b)). 
One readily finds that
\begin{equation}
\label{Ef}
V_f\simeq 2E_c\sqrt{lt_W}\left(1-0.6\sqrt{h/l}\right).
\end{equation} 
Here $E_c$ is the thermodynamic coercive field,  which 
was evaluated as \cite{tagantsev2010domains}
\begin{equation}
\label{Ec}
E_c\simeq\frac{0.2 P_0}{\chi}.
\end{equation} 

Equation (\ref{Ef}) gives the critical voltage $V_f$ in terms of fundamental parameters of the ferroelectric and the dimensions of the sample.
However, to evaluate $V_f$,  it is more practical to use the surface energy of ferroelastic $90^\circ$ wall $\sigma_f$ and the effective isotropic dielectric susceptibility $\chi$.
Starting from  (\ref{Us0}), (\ref{UD}), (\ref{condc}), and (\ref{UE}), we readily find the following alternative form of Eq.~(\ref{Ef})
\begin{equation}
\label{Ef1}
V_f\simeq \sqrt{\frac{\sigma_f l}{\chi}}\left(1-0.6\sqrt{h/l}\right).
\end{equation} 

To numerically evaluate $V_f$, we set $\sigma_f= 30\, m\mathrm{J}/\mathrm{m}^2$ and  $\chi=250\cdot8.85\cdot10^{-12} \frac{C}{V\cdot\mathrm{m}}$.

The theoretical estimates for $\sigma_f$ were taken from Ref.~\cite{tagantsev2010domains,kinase1970on}, and the effective dielectric susceptibility was considered as the in-series combination of two tensor components of the susceptibility of BaTiO$_3$ at room temperature \cite{remeika1954method}.
Using the above setting, for $l= 5\cdot10^{-3}\mathrm{~m}$ and $h= 1\cdot10^{-3}\mathrm{~m}$, Eq. (\ref{Ef1}) yields $V_f\simeq~200\,\mathrm{~V}$, which nicely agrees with the experiment.
Note that in the experiment the threshold voltage may be higher than the theoretical prediction in view of the pinning of domain walls on lattice and defects.
Dissipative forces and barriers prevent the transition exactly at the moment when the energies of the two states are equalized and delay it until the energy of the final state becomes essentially lower.
It follows from Eq. (\ref{Ef}) that  the  critical voltage $V_f$ scales as the square root of the sample size if the aspect ratio of the length to thickness is preserved.

\section{Phase field simulations}

The obtained scaling law allow modeling the phenomenon on smaller samples.
This is vital in view of the necessity to resolve individual domain walls of a nanometer size.
No simulations for a millimeter sample would be possible because of computational power constraints.
In the calculations, we took 100 nm  thick sample, and  kept the aspect ratio as in the experiment.
Then the results of the simulations were analyzed using the scaling law.
The model solves Landau-Ginsburg-Devonshire (LGD) equations, equations of elasticity and electrostatics in a 2D statement, see details in Appendix, section \ref{PhFM}.

\begin{figure}[t]
\centering
\includegraphics{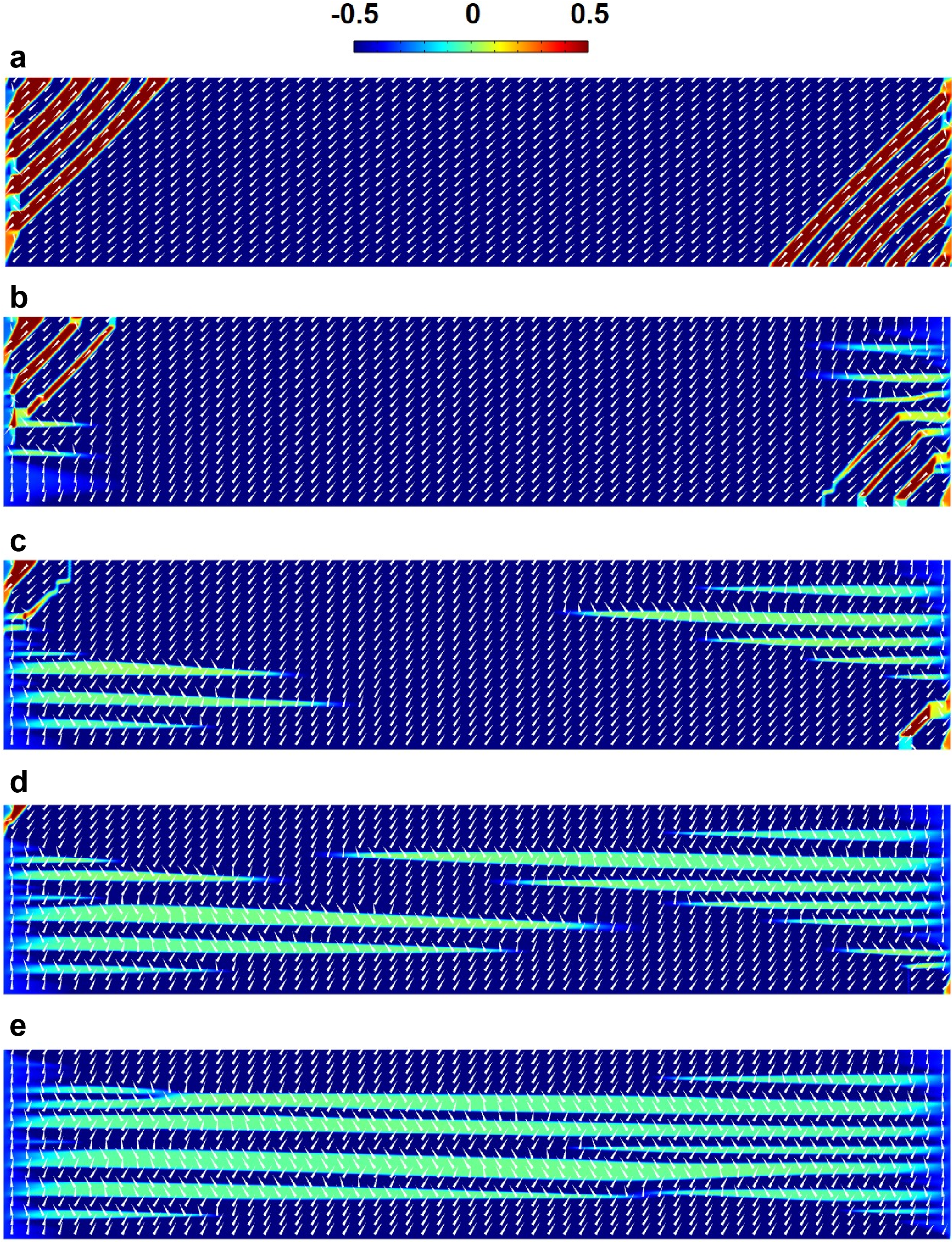}
\caption{
Numerical simulations of domain pattern of BaTiO$_3$ under applied electric voltage $V$  along [110] direction.
(a)~-~relaxed structure at $V$~=~0~V, (b)~-~$V$~=~5~V, (c)~-~$V$~=~10~V, (d)~-~$V$~=~15~V, (e)~-~$V$~=~20~V.
Color map is provided for the value of $-1.5P_x-P_y$. Arrows show polarization vector.
}
\label{Carrots}
\end{figure}

The simulations qualitatively reproduce the experimental observation, where a transition from a mostly-monodomain state to a polydomain state with a dense array of horizontal ferroelastic walls takes place.
Fig.\,\ref{Carrots} shows frames of a simulation where voltage was set to 0~V and after full relaxation (a) slowly increased to 5~V~(b), 10~V~(c) and further to 15~V~(d) and 20~V~(e) in  a sample with height $h~=100$~nm and length $l~=500$~nm.
Initial conditions for polarization were set exactly in accordance with the Fig. \ref{Theory}~b. The initial polarization distribution in the absence of applied voltage relaxed to a state shown in Fig.  \ref{Carrots}~a, where one may notice some perturbations of the domain stripes near the edges.
When the electric field was applied, the stripes where polarization projected oppositely to the applied electric field became energetically unfavorable and shrunk: First the longest, Fig.\,\ref{Carrots} b, then the shorter ones, Fig.\,\ref{Carrots}~c,~d.

Thus the pattern of the $180^\circ$ walls act as a shutter that is removed by the electric field making way for the growth of the carrot-shaped domains.
Above the critical voltage the "carrots" growing from the two sides of the sample "shake hands" resulting in a domain structure corresponding to the one shown in Fig.\,\ref{Theory} c, which we directly describe using the classical theory of Mitsui and Furuichi\cite{mitsui1953domain}.

In the simulations we achieved a gradual transition from the quasi-monodomain to polydomain structure in the range of voltage from 5 to 20~V for 100~nm thick and 500~nm long sample.
According to the scaling law one recalculates $V_f$ as 500~V~-~2~kV for a sample with thickness 1~mm and length 5~mm, which roughly agrees with the experiment, where the transition took place from 200 to 600~V.
The slight disparity may be explained by stronger pinning of domain walls on the computational mesh than in real conditions.
Also the effects caused by  semiconducting properties of the sample are not taken into account in the model.
\begin{figure}[h!]
\centering
\includegraphics{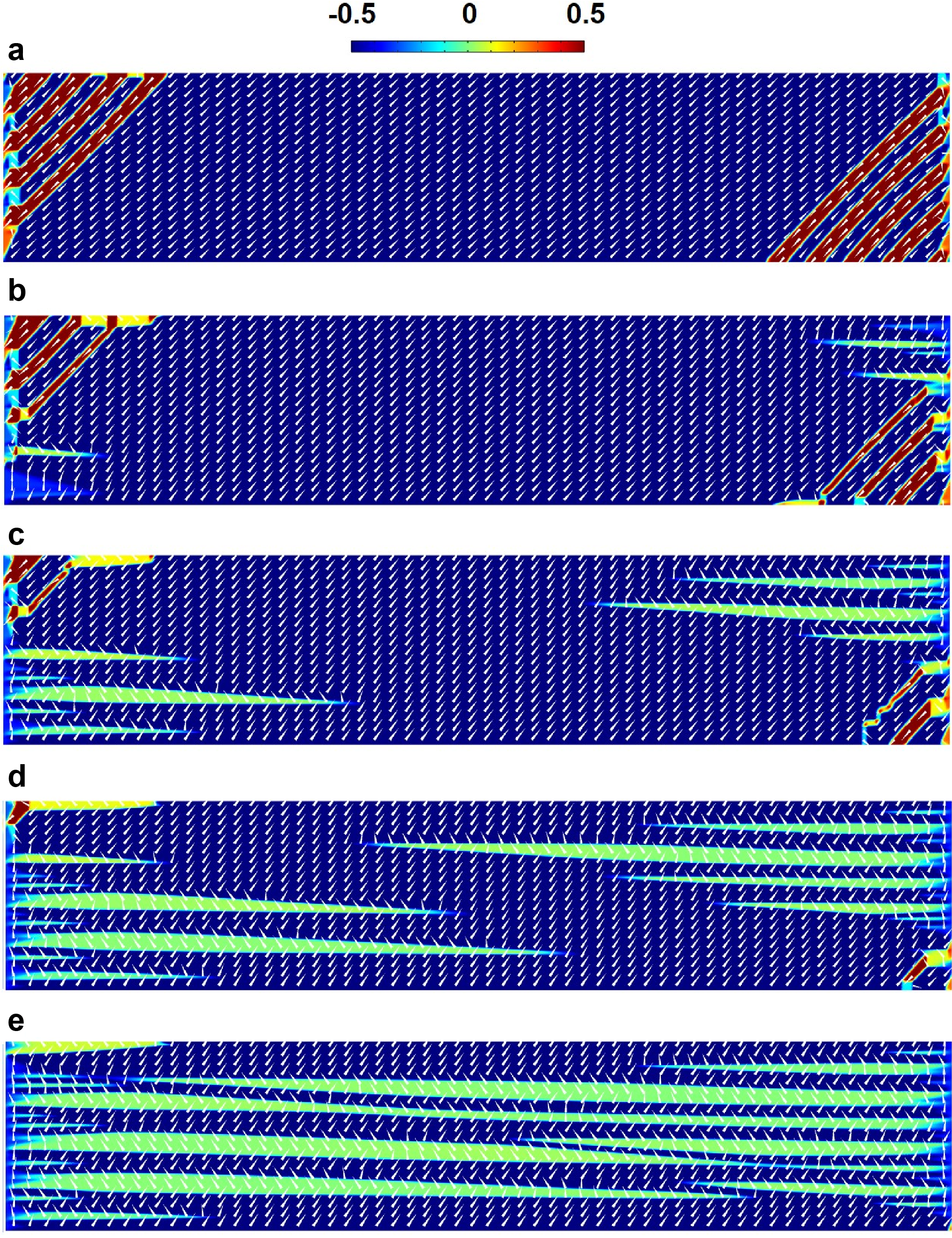}
\caption{
Numerical simulations of domain pattern of BaTiO$_3$ under applied electric voltage $V$  with electromechanical coupling disabled.
(a)~-~relaxed structure at $V$~=~0, (b)~-~$V$~=~5~V, (c)~-~$V$~=~10~V, (d)~-~$V$~=~15~V, (e)~-~$V$~=~20~V.
Color map is provided for the value of $-1.5P_x-P_y$.
Arrows show polarization vector.
}
\label{NoElast}
\end{figure}
\begin{figure}[h!]
\centering
\includegraphics{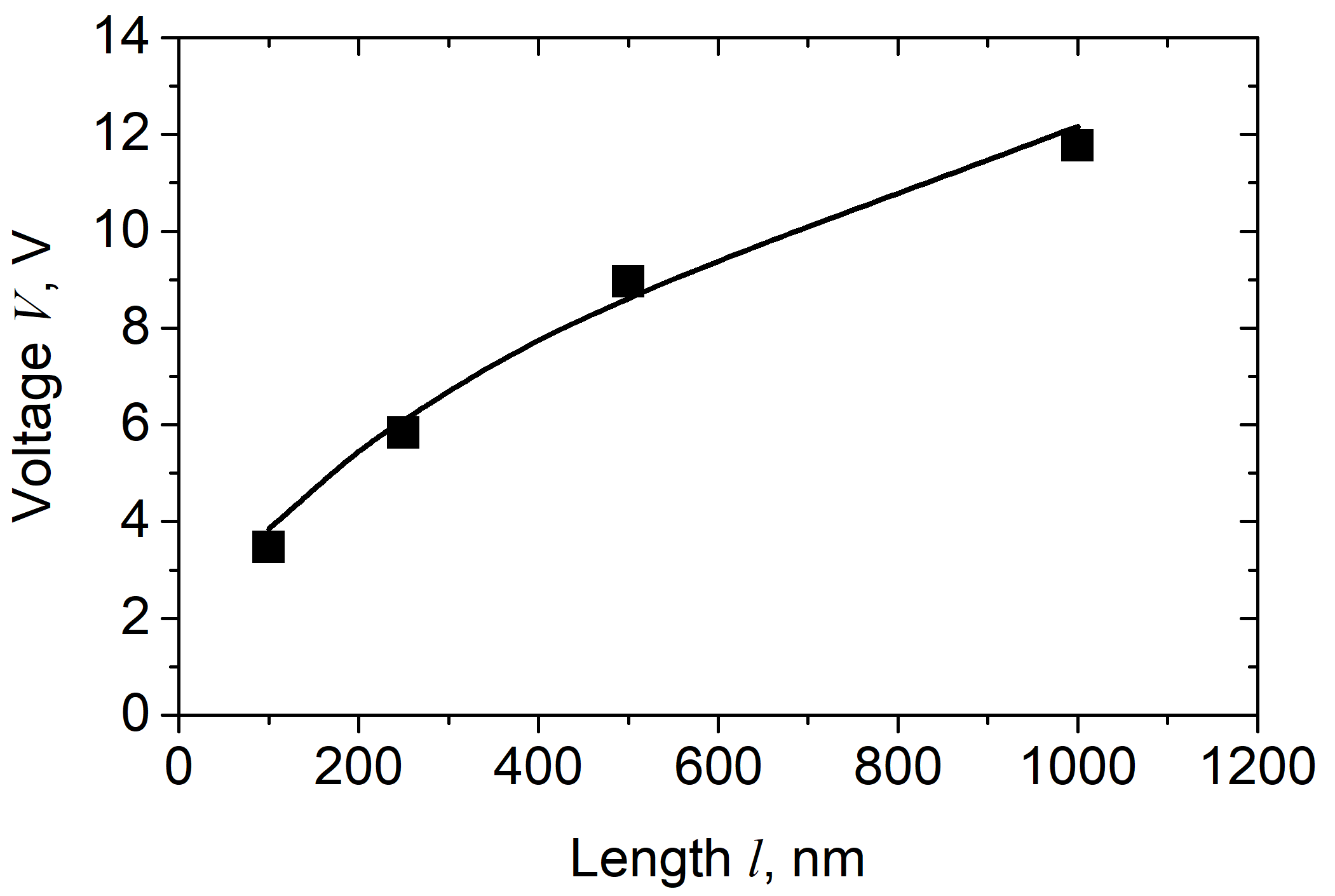}
\caption{
Numerical validation of the scaling law. The simulated critical voltage (squares) for samples with the dimensions ($h\times l$): $20\times100$~nm$^2$ ; $50\times250$~nm$^2$; $100\times500$~nm$^2$; $200\times1000$~nm$^2$ and the square root  theoretical fit by Eq.~(\ref{Ef}) with constant aspect ratio $l=5h$~(\ref{Ef}).
In simulations the critical voltage corresponds to the moment when the first domain stripe grows through the entire sample.}
\label{ScalingLaw}
\end{figure}

\begin{figure}[h!]
\centering
\includegraphics{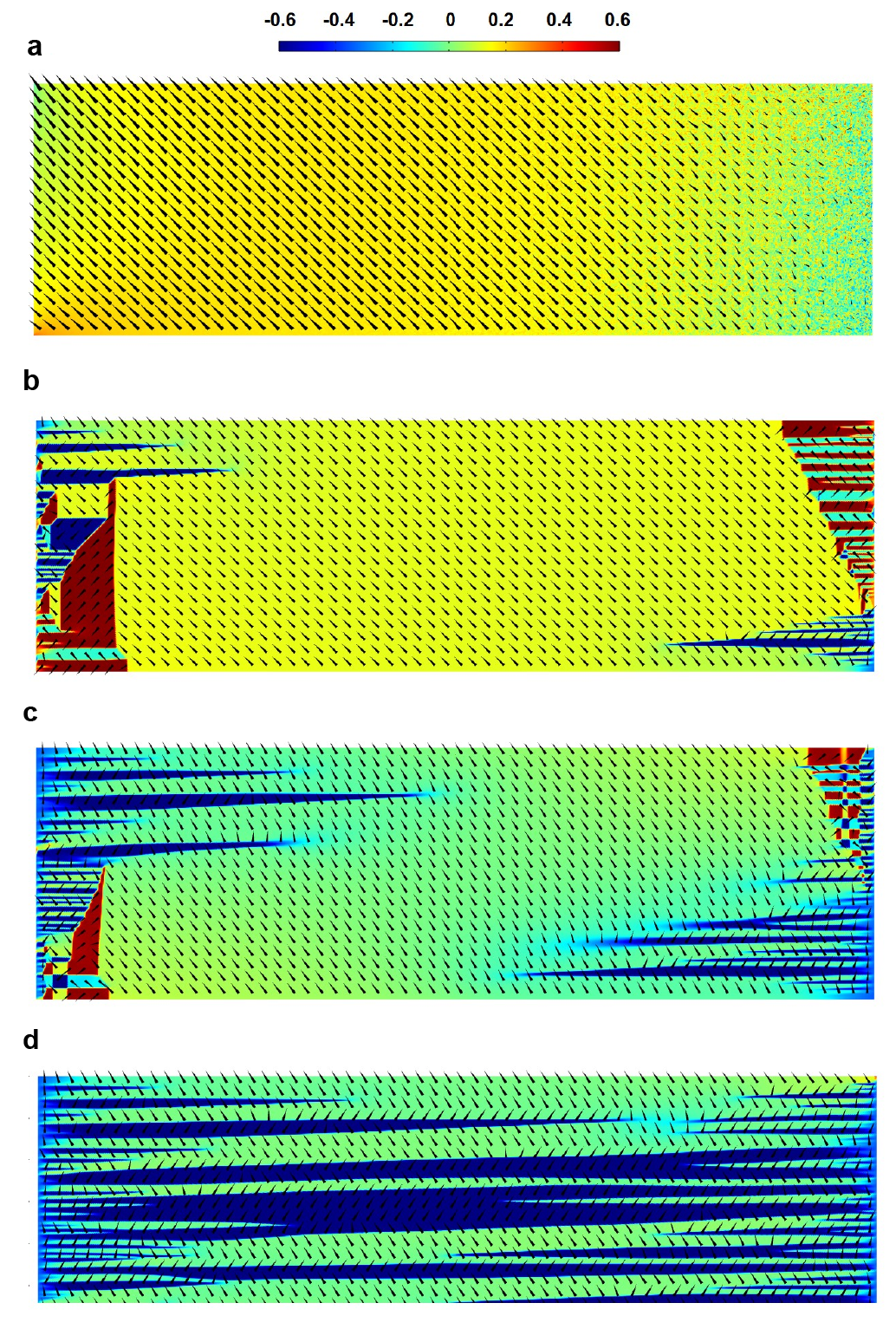}
\caption{
Numerical simulations of domain pattern of BaTiO$_3$ under applied electric voltage $V$  along [110] direction in a 300~nm by 1000~nm sample.
(a)~-~initial conditions for polarization used in accordance with Appendix, Eq. (\ref{Pinit}), (b)~-~relaxed structure at $V$~=~0~V, (c)~-~$V$~=~25~V, (d)~-~$V$~=~50~V.
Color map is provided for the value of $-2P_x-P_y$.
Arrows show polarization vector.
}
\label{SimulSup}
\end{figure}

In support to the above considerations we performed additional simulations. 
First, we repeated the calculations leading to the domain pattern evolution shown in Fig.\,\ref{Carrots} but with the mechanical effects being turned off  by setting all the electrostrictive  coefficients to zero.
The results of these calculations are shown in Fig.\,\ref{NoElast}.
It is seen that the results obtained  with and without electrostriction differ only slightly.
This fact confirms the validity of the analytical model where only electrostatic effects are considered.
Elastic effects lead to straightening of the wedge domains, such that they better resemble  experimentally observed stripes.

Further we illustrated that our results do not rely on the choice of the specific initial domain structure by performing a series of simulations  with different other initial conditions on polarization.
The results of the simulation are illustrated  in Fig.\,\ref{SimulSup}.
Comparing Fig.\,\ref{SimulSup} with Fig.\,\ref{Carrots} we see that for the two sets of completely different initial conditions and accordingly distinct domain patterns near the edges, the settlement of the striped pattern is astonishingly similar.
Thus our choice of the initial structure with 180-degree domain walls which was consistent but not verified by experimental observations, is appropriate.

Finally, we numerically verified the scaling law, Eq.~(\ref{Ef}), by tracing the dependence of the critical voltage on the sample size (Fig.\,\ref{ScalingLaw}). The results for the  samples with the dimensions ($h\times l$) of $20\times100$~nm$^2$, $50\times250$~nm$^2$, $100\times500$~nm$^2$ and $200\times1000$~nm$^2$ are in good agreement with the square root  theoretical fit by Eq. (\ref{Ef}) with constant aspect ratio $l=5h$.  

\section{Conclusions}

We experimentally demonstrated a paradoxical situation where the electric field induces stripe domains of a new domain state into an initial single-domain crystal even though the applied field favors both final domain states equally.
This experimental finding was rationalized using an analytical model and numerical phase field simulations.
The phenomenon was understood as a result of an interplay between the domain wall energy and the energy of the depolarizing field, which, in our system, due to finite size effects and the geometry of the sample, becomes very specific.
The results of the paper can offer new possibilities for domain engineering.

\section{Data Availability Statement}

All data generated or analyzed during this study are included in this article and available from the authors upon reasonable request.

\section{Acknowledgements}

The study was conceived and carried out with the support from the Czech Science Foundation (GACR project No.20-05167Y).
Authors acknowledge the assistance provided by the Ferroic Multifunctionalities project, supported by the Ministry of Education, Youth, and Sports of the Czech Republic. Project No.CZ.02.01.01/00/22$\underline{\,\,\,}$008/0004591, co-funded by the European Union.
%


%
%
%

\clearpage
\newpage

\appendix{Appendix}


\section{Experimental technique}\label{Methods}
The $\langle 110 \rangle$ oriented barium titanate (BaTiO$_3$) single crystals were obtained through the top-seeded solution growth technique (TSSG) from Electro-Optics Technology GmbH.
The samples were shaped as bars with dimensions of 1x1x5 mm$^3$, aligned with the longest edge along the $[\bar{1}10]$ direction. 
The (001) plane was polished to a quality of 1\textmu m, and the domain structure was observed along the [001] direction.
Gold electrodes were applied to the (110) planes, and an electric field was applied along the [110] direction using a high voltage DC power supply (SRS PS325).
Contacts to the electroded surface were made using high-temperature silver paste.
Samples with various domain structures, including the monodomain state and state with (110) domain walls, were examined at room temperature using a polarizing microscope (Leica DM2700M) in both transmission and reflection modes.
The process of domain wall formation was documented using software provided by Leica (LAS X).

For our experiments we selected samples that were initially free of (110) ferroelastic domain walls.
We poled the sample by gradually applying an electric field along the [110] direction.
This process either removed the minority of domains with unfavorable polarization or switched the entire crystal, depending on its initial polarity.
The primary switching process occurred at an electric field of approximately 1.3~kV/cm, indicating that fields above this threshold effectively eliminate domains with opposite polarization directions.
In both cases, the resulting state was a monodomain poled state.
After the removal of the electric field, no ferroelastic domain walls were observed; however, some  nonferroelastic 180-degree domain walls may be present at the edges of the sample.
The formation of a through-the-crystal polydomain pattern was observed during a similar poling procedure, for the  fields from 3 to 6~kV/cm.

\section{Implications of Mitsui's theory} \label{Mitsui}

\begin{figure}[b]
\centering
\includegraphics{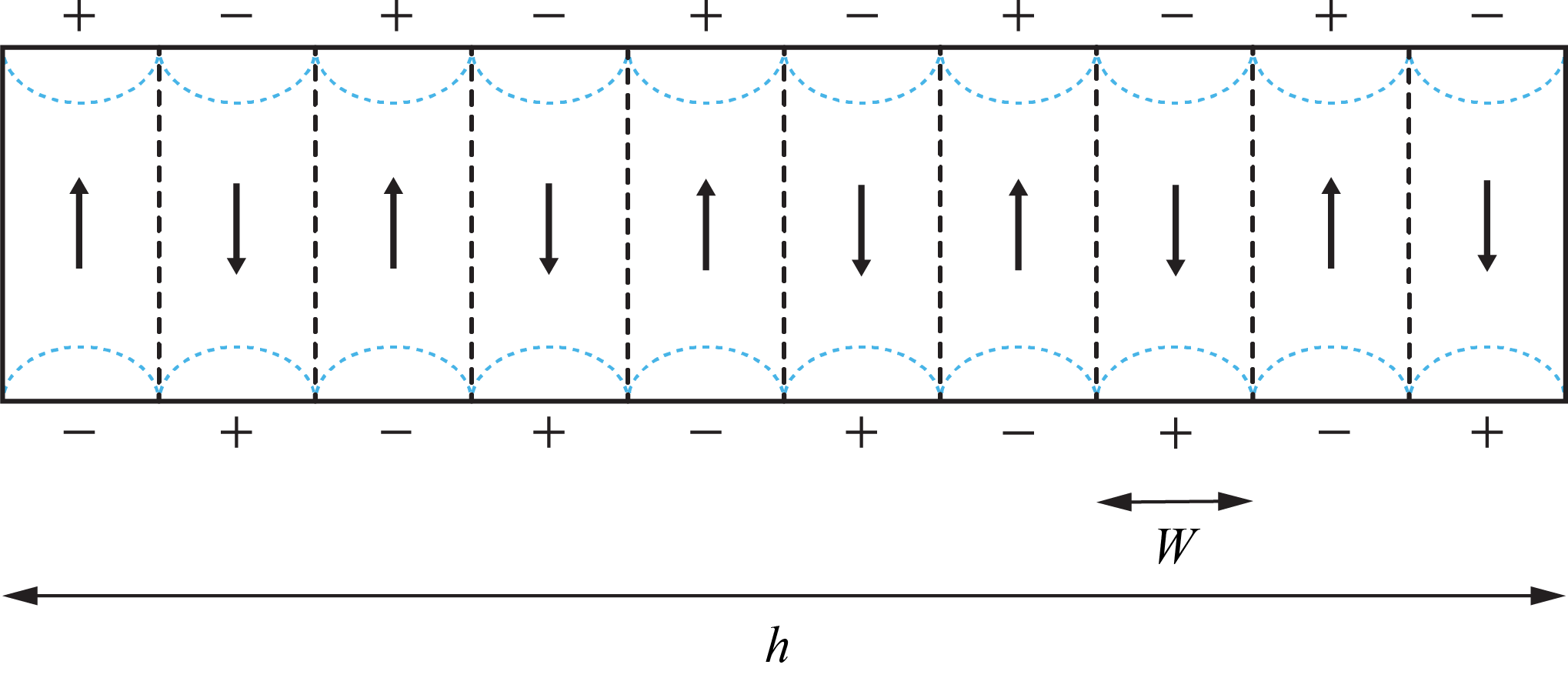}
\caption{
The domain pattern in a ferroelectric plate described by  theory of Mitsui \cite{tagantsev2010domains, mitsui1953domain}.
Dashed lines show $180^\circ$ domain walls.
Arrows show the orientation of the spontaneous polarization.
The sign of the total (bound + free) surface charge is indicated.
The regions with strong depolarising field are schematically outlined with thin dashed lines.
}
\label{TheorySup}
\end{figure}

It is known that in a ferroelectric plate with free (non-electroded) surfaces, a single-domain state is not thermodynamically stable due to the depolarizing electric fields from bound charges present at the surfaces.
For this reason the ferroelectric sample splits into domains, where depolarizing fields penetrate into the sample only at the distance comparable to the domain width, at longer distances fields of different sign compensating each other.
This is schematically illustrated in Fig.\,\ref{TheorySup}.
Mitsui and Furuichi obtained analytical solution to the associated electrostatic problem, and obtained a scaling law for the dependence of domain width as a function of sample thickness, which is analogous to the Kittel's law for ferromagnets.
Mitsui's solution was obtained in the hard ferroelectric approximation where the vector of polarization  $P_i$ is presented  as a sum of the spontaneous $(P_0)_i$ contribution and that which is linear in the  electric field $E_i$:
\begin{equation}
\label{hard}
P_i = (P_0)_i +\chi _{ij}E_j ,
\end{equation} 
where the suffices numerate the Cartesian components and the dummy suffix summation rule is adopted.
In Ref. \cite{mitsui1953domain}, the principle axes of the dielectric susceptibility $\chi_{ij}$ were assumed to be either parallel or normal to the surface.
It is seen that the periodical system of stripes of the bound charge on the lateral sides of the beam in the case shown in Fig.\,\ref{Theory} c is geometrically identical to that in the Mitsui's problem.
At the same time, one readily checks that, for the case illustrated in this figure, for tetragonal BaTiO$_3$, the principle axes of  $\chi_{ij}$ are obliquely oriented with respect to the plane of the plate.
Keeping in mind a semi-quantitative model, we neglect the anisotropy of $\chi_{ij}$, i.e. we set $\chi_{ij}= \chi\delta_{ij}$.
In this case, we can use Mitsui's solution for the energy of the depolarizing field once the domain pattern shown in 
Fig.\,\ref{Theory} c is dense, i.e.
\begin{equation}
\label{cond01}
W\ll h,
\end{equation} 
where $W$ is the distance between the adjacent walls . 

Using Mitsui's solution for the energy of the depolarizing field we can directly evaluate the free energy of the configuration shown in Fig. \ref{Theory} c, utilizing the geometric similarity.
A two-dimensional model is sufficient; accordingly, the energy of the system is calculated per unit length in the third dimension.
The combined assumptions of Mitsui's model and our assumptions are as follows: (i) the ferroelectric is addressed in the so-called  hard ferroelectric approximation \cite{tagantsev2010domains}, (ii) For the estimates for the thickness and energy of both $90^\circ$ and $180^\circ$ walls, we use the results for a $180^\circ$ wall in the case of the second order phase transition, (iii) The anisotropy of dielectric permittivity of the ferroelectric is neglected.  

Under the above conditions and approximations we arrive at the  expression for the energy of the depolarizing field  \cite{mitsui1953domain,tagantsev2010domains},
\begin{equation}
\label{EsS}
\mathcal{E}_c = \frac{0.13 P_0^2 W}{\chi}h,
\end{equation} 
where it is taken into account that the dielectric permittivity is much larger than that of the free space and that, in our case, the surface density of the bound charge  at the lateral sides  equals $ P_0\sqrt{2}/2 $, where $P_0 $ is the spontaneous polarization.
In the main text we rewrite Eq. (\ref{EsS}) using Eq. (\ref{U0}) to get Eq. (\ref{UD}).
While application of Mitsui's theory to domain structures shown in Fig.\,\ref{Theory} c is geometrically straightforward, we can also apply it to those shown in Fig.\,\ref{Theory} b. This is done by neglecting the distortions of the depolarizing field caused by oblique contacts of domain walls with the side surface.
Taking into account that now only one end of the pattern contributes to the energy of the depolarizing field and that the average length of the walls in the pattern is equal to $h/\sqrt{2}$, the optimized energy of the system $U_b$ reads:
\begin{equation}
\label{UbS}
U_b \approx U_0\sqrt{ht_W}h.
\end{equation} 
When writing (\ref{UbS}) (Eq. (\ref{Ub}) of the main text) we  assume that the pattern is periodic.
It occurs that, for the single-domain configuration shown in Fig.\,\ref{Theory} a, Eq. (\ref{EsS}) yields exactly its energy $U_a$ if $W$ is replaced with $h$ to yield:
\begin{equation}
\label{UaS}
U_a \approx 0.4 U_0h^2.
\end{equation} 
This follows from the method of mirror images for the electric fields in the presence of electrodes \cite{LANDAU198434}.

\section{The phase field model} \label{PhFM}
Model equations are obtained by Lagrange principle from Helmholtz free energy density \cite{li2006temperature}:
\begin{eqnarray}\label{LGD}
f[\{P_i,P_{i,j},e_{ij},D_i\}]=f_{\text{bulk}}^{(e)}+f_{\text{ela}}+f_{\text{es}}+f_{\text{grad}}+f_{\text{ele}},
\end{eqnarray}
where $P_i$ is the ferroelectric part of polarization, $P_{i,j}$ its derivatives (the subscript '${,i}$' represents the operator of spatial derivatives $\partial/\partial x_i$), $D_i$ the electric displacement and $e_{ij}=1/2(u_{i,j}+u_{j,i})$ is the elastic strain where $u_i$ is a displacement vector.

The bulk free energy density
{\small \begin{eqnarray}\label{fbulk}
\nonumber \lefteqn{f_{\text{bulk}}^{(e)}[\{P_i\}]=}\\
\nonumber  &&\alpha_{1}\displaystyle\sum_i P_i^2+\alpha_{11}^{(e)}\displaystyle\sum_iP_i^4+\alpha_{12}^{(e)}	 \displaystyle\sum_{i>j}P_i^2 P_j^2+\alpha_{111}	\displaystyle\sum_{i}P_i^6\\
 &&+\alpha_{112}	\displaystyle\sum_{i>j}(P_i^4 P_j^2+P_j^4 P_i^2)+\alpha_{123}	 \displaystyle\prod_{i} P_i^2
\end{eqnarray}}
is expressed for a zero strain as a six-order polynomial expansion \cite{hlinka2009piezoelectric}, where $\alpha_i,
\alpha_{ij}^{(e)}, \alpha_{ijk}$ are parameters fitted to the single crystal properties (Table \ref{tab:par}).
The remaining contributions represent bilinear forms of densities of elastic energy
$f_{\text{ela}}[\{e_{ij}\}]=1/2 c_{ijkl} e_{ij} e_{kl}$, where $c_{ijkl}$ is the elastic stiffness,
electrostriction energy $f_{\text{es}}[\{P_i,e_{ij}\}]=-q_{ijkl} e_{ij} P_k P_l$, where $q_{ijkl}$ are the electrostriction coefficients, gradient energy $f_{\text{wall}}[\{P_{i,j}\}]=1/2 G_{ijkl} P_{i,j} P_{k,l}$, where $G_{ijkl}$ are the gradient energy coefficients, and electrostatic energy $f_{\text{ele}}[\{P_i,D_i\}]=1/(2 \varepsilon_0 \varepsilon_B) (D_i-P_i)^2$, where
$\varepsilon_0$ and $\varepsilon_B$ are permittivity of vacuum and relative background permittivity, respectively.
The zero-strain coefficients $\alpha_{ij}^{(e)}$ can be expressed in terms of usually introduced stress-free coefficients $\alpha_{ij}$ as follows:
\begin{eqnarray*}
\label{alpha11}
  \alpha_{11}^{(e)} &=& \alpha_{11}+\frac{1}{6} \left(\frac{2(q_{11}-q_{12})^2}{c_{11}-c_{12}}+ \frac{(q_{11}+2 q_{12})^2}{c_{11}+2 c_{12}}\right), \\
\label{alpha12}
  \alpha_{12}^{(e)} &=& \alpha_{12}+\frac{1}{6} \left(\frac{2(q_{11}+2 q_{12})^2}{c_{11}+2 c_{12}}-\frac{2(q_{11}-q_{12})^2}{c_{11}-c_{12}}+\frac{3 q_{44}^2}{4 c_{44}}\right).
\end{eqnarray*}

By using the Legendre transformation to the thermodynamic potential with a fixed voltage at the electrodes
\begin{eqnarray}\label{enthalp}
\nonumber
h[\{P_i,P_{i,j},u_{i,j},\varphi_{,i}\}] = f[\{P_i,P_{i,j},e_{ij},D_i\}]-D_i
E_i,
\end{eqnarray}
where $E_i=-\varphi_{,i}$ is the electric field and $\varphi$
is the electric potential, and using Lagrange principle, we can uniformly express the set of field equations which govern the kinetics of ferroelectrics:
\begin{eqnarray}\label{eqs}
\label{mech}\left(\frac{\partial h}{\partial e_{ij}}\right)_{,j}&=&0,\\
\label{diel}\left(\frac{\partial h}{\partial E_{i}}\right)_{,i}&=&0,\\
\label{GD}\frac{1}{\Gamma}\frac{\partial P_{i}}{\partial t}-\left(\frac{\partial h}{\partial P_{i,j}}\right)_{,j}&=&-\frac{\partial h}{\partial P_{i}}.
\end{eqnarray}
Equation (\ref{mech}) defines the mechanical equilibrium while inertia is neglected.
The Poisson's Eq. (\ref{diel}) represents Gauss's law for charge and electric field in a dielectric.

The model geometry is a block of [110] oriented BaTiO$_3$ crystal with thickness $h=300$~nm  and length $l=1\mu$m, except for some simulations where the other size is specified.

Mechanically free boundary conditions were used for all surfaces.
The bottom and top surfaces had the electric potential fixed to $\varphi=0$ and $\varphi=V(t)$, respectively, where $V(t)=0$ if $0<t<t_0$ and $V(t)=V_{max}(t-t_0)/(t_{max}-t_0)$ if $t>t_0$.
Here $V_{max}$ is the maximum applied voltage, $t_{max}$ is the simulation time set in the model; $t_0$ and $t_{max}$ were set large enough to ensure convergence of the solution.
The condition of zero free charge was set on the side surfaces, if not indicated otherwise.
The initial conditions for polarization were set as for [010]-oriented monodomain, except for some simulations where other condition is specified.
The initial electric potential in the film is zero and the initial mechanical displacement corresponds to a stress-free substrate in the whole model.
The model is numerically solved by the finite element method with a time dependent solver in COMSOL 5.3.
Variations of simulation parameters include aspect ratio, initial condition for polarization, simulations with partly screened side surfaces, and a test with elastic energy excluded.

\begin{table}[t]
  \centering
\begin{tabular}{|c|c|c|}
  \hline
  Parameter & Value & Unit \\
  \hline
  $\alpha_1$ & $(T-381) 3.34 \times10^{5}$ & $Jm/C^2$  \\
  $\alpha_{11}$ & $(T-393) 4.69 \times10^{6} -2.02\times10^8$ & $Jm^5/C^4$  \\
  $\alpha_{12}$ & $3.23\times10^8$ & $Jm^5/C^4$ \\
  $\alpha_{111}$ & $-(T-393) 5.52 \times10^{7} +2.76\times10^9$ & $Jm^9/C^6$ \\
  $\alpha_{112}$ & $4.47\times10^9$ & $Jm^9/C^6$  \\
  $\alpha_{123}$ & $4.91\times10^9$ & $Jm^9/C^6$ \\
  \hline
  $c_{11}$ & $27.5\times 10^{10}$ & $J/m^3$ \\
  $c_{12}$ & $17.9\times10^{10}$ & $J/m^3$ \\
  $c_{44}$ & $5.43\times 10^{10}$ & $J/m^3$ \\
  \hline
  $q_{11}$ & $14.2\times10^{9}$ & $Jm/C^2$  \\
  $q_{12}$ & $-0.74\times10^{9}$ & $Jm/C^2$ \\
  $q_{44}$ & $6.28\times10^{9}$ & $Jm/C^2$ \\
  \hline
  $G_{11}$ & $51\times10^{-11}$ & $Jm^3/C^2$ \\
  $G_{12}$ & $-2\times10^{-11}$ & $Jm^3/C^2$ \\
  $G_{44}$ & $2\times10^{-11}\times10^{-10}$ & $Jm^3/C^2$ \\
  \hline
  $\Gamma$ & $4\times10^4$ & $C^2/(Jms)$ \\
  \hline
  $\varepsilon_B$ & $10$ & $1$  \\
  \hline

\end{tabular}
\caption{Values of material coefficients for BaTiO$_3$ used in the phase field simulations, Ref. \cite{hlinka2009piezoelectric}}\label{tab:par}
\end{table}

\section{Supplementary simulation details} \label{SimulSuppl}

We performed a series of numerical simulations with different initial conditions on polarization to illustrate that our results do not rely on the hypothesis of a specific initial structure with 180-degree domain walls being stable with respect to these changes.
Figure \ref{SimulSup} shows frames of a simulation where voltage was set to 0 (a) and after full relaxation (b) linearly increased to 30V (c) and further to 50 V (d) in  a sample with height $h =300$ nm and length $l =1 \mu$m.
Here in order to highlight that the result is general we use different initial conditions for polarization near the two edges of the sample:
\begin{equation}
\label{Pinit}
P_{init}=\frac{P_0}{\sqrt{2}}\left(\binom{-1}{1}-2\exp(3x/2h)\binom{Rn1(x,y)}{Rn2(x,y)}  \right),
\end{equation} 
where $Rn1$ and $Rn2$ are random functions with uniform distribution from 0 to 1, $x$ is measured from the left sample edge.
This corresponds to a nearly pure monodomain state on the left-hand side and a random polarization on the right hand side, see Fig.\,\ref{SimulSup}~a for initial conditions on polarization.
This polarization distribution in the absence of applied voltage relaxed to a state shown on Fig.\,\ref{SimulSup}~b, which is monodomain in the middle and polydomain near the edges.
Common feature of the ferroelastic polydomain patterns near the edges is that they contain an area of domains with inverted vertical polarization component to ensure flux closure. 
%


\end{document}